\input iopppt.modifie
%
\def\received#1{\insertspace 
     \parindent=\secindent\ifppt\textfonts\else\smallfonts\fi 
     \hang{Received #1}\rm } 
\headline={\ifodd\pageno{\ifnum\pageno=\firstpage\titlehead
   \else\rrhead\fi}\else\lrhead\fi} 
\def\endtable{\parindent=\textind\textfonts\rm\bigskip} 
\def\rrhead{\textfonts\hskip\secindent\it 
    \shorttitle\hfill\rm L\folio} 
\def\lrhead{\textfonts\hbox to\secindent{\rm L\folio\hss}%
   \it\aunames\hss} 
\footline={\ifnum\pageno=\firstpage
\hfil\textfonts\rm L\folio\fi}   
\def\titlehead{\smallfonts J. Phys. A: Math. Gen. {\bf 19} (1986)
L411--L414 
\hfil} 

\firstpage=411
\pageno=411


\jnlstyle

\jl{1}

\letter{Universal amplitude of the free energy density\hfill\break in
finite-size scaling: the Potts universality}

\author{L Turban and J-M Debierre}
\address{Laboratoire de Physique du 
Solide\footnote{$\dagger$}{Laboratoire associ\'e au CNRS no 155.}, Ecole des
Mines, Parc de Saurupt, F54042 Nancy cedex and Universit\'e de Nancy I,
BP~239,  F54506~Vand\oe uvre l\`es Nancy, France}

\received{27 January 1986}

\abs
Using the numerical results of the finite-size scaling study of the $q$-state
Potts model by Bl\"ote and Nightingale, we obtain the following conjectured
expressions for the universal amplitude of the free energy density:
$$
A_0(u)=\pi\,{(2-3u)(u+1)\over6(2-u)}
$$
where
$$
0\leq u\equiv{2\over\pi}\,\cos^{-1}\!\left({\sqrt{q}\over2}\right)\leq1\,.
$$
\endabs

\vglue1cm
According to Privman and Fisher (1984), the singular part of the free energy
density on a cylinder-shaped system with size $V=L^{d-1}\times\infty$ near the
critical point $t=0$, $h=0$ ($t=(T-T_{\rm c})/T_{\rm c}$, $h=H/k_{\rm B}T$)
may be written as
$$\fl
f^{(s)}(t,h,L)=-{F^{(s)}\over Vk_{\rm B}T}\simeq L^{-d}Y(x_1,x_2)\qquad
x_1=C_1tL^{y_t}\qquad x_2=C_2hL^{y_h}
\eqno(1)
$$
where $Y(x,y)$ is a universal function, $y_t$ and $y_h$ are the thermal and
magnetic exponents. $C_1$ and $C_2$ are non-universal metric factors.

In two dimensions, on a $L\times\infty$ strip with periodic boundary
conditions built up of $L\times l$ slices, the free energy density is given
by
$$
f_0(t,h,L)=(1/lL)\ln\Lambda_0(t,h,L)
\eqno(2)
$$
where $\Lambda_0$ is the largest eigenvalue of the transfer matrix. Free
energy levels $f_j(t,h,L)$ may be defined using the subdominant eigenvalues
$\Lambda_j$ of the transfer matrix ($\Lambda_0>\Lambda_1\geq\Lambda_2\cdots$)
through
$$
f_j(t,h,L)=(1/lL)\ln\Lambda_j(t,h,L)\,.
\eqno(3)
$$
The singular part of the free energy levels is expected to behave as in
equation (1):
$$
f^{(s)}_j(t,h,L)\simeq L^{-2}Y_j(x_1,x_2)\,.
\eqno(4)
$$
The singular part is defined as
$$
f_j(t,h,L)=f^{(s)}_j(t,h,L)+f_\infty(t,h)
\eqno(5)
$$
where the last term is the analytic background which is the same for all the
levels.

At the critical point one obtains
$$
f_j=A_jL^{-2}+B
\eqno(6)
$$
where $A_j=Y_j(0,0)$ is a universal amplitude and $B$ is the critical value of
the free energy density in the infinite system.

The correlation lengths are given by
$$
\xi_{\parallel j}(t,h,L)
=l\left[\ln\left({\Lambda_0\over\Lambda_j}\right)\right]^{-1}
\eqno(7)
$$
($j\!=\!1$: spin-spin correlations, $j\!=\!2$: energy-energy correlations) so
that, using equations (3) and (6), one obtains
$$
\xi_{\parallel j}(0,0,L)=S_jL
\eqno(8)
$$
where
$$
S_j=(A_0-A_j)^{-1}\,.
\eqno(9)
$$
These correlation length universal amplitudes $S_j$ are known to be related to
the decay exponents $\eta_j$ (Pichard and Sarma 1981, Luck 1982, Derrida and
de S\`eze 1982, Nightingale and Bl\"ote 1983, Cardy 1984) through
$$
S_j=1/\pi\eta_j
\eqno(10)
$$
so that
$$
A_0-A_j=\pi\eta_j\,.
\eqno(11)
$$
\midinsert
\table{Free energy density at the critical point in the $q$-state Potts model as a
function of the strip width $L$ (data taken from Bl\"ote and Nightingale 1982). The
values converge towards the exact result $B$ (Baxter 1973) for the infinite
 system.}[w]
\align\L{#}&&\L{#}\cr  
\br  
&\centre{4}{$L$}\cr  
\ns\ns
&\crule{4}\cr
$q$&$\ \;\,9$&$\ \;\,10$&$\ \;\,11$&$\ \;\,\infty$\cr
\mr\cr
\ns\ns\ns
1/64&$-0.893\ 297\ 995\ 296$&$-0.891\ 242\ 331\ 139$&$-0.889\ 723\ 754\ 951$
&$-0.882\ 519\ 177\ 979$\cr  
1/16&$-0.168\ 550\ 388\ 717$&$-0.166\ 874\ 332\ 373$&$-0.165\ 636\ 316\ 196$
&$-0.159\ 764\ 272\ 049$\cr 
1/2&$\ \; \,0.977\ 625\ 265\ 191$&$\ \; \,0.978\ 178\ 682\ 825$&$\ \; \,0.978\ 587\ 322\ 901$
&$\ \; \,0.980\ 523\ 980\ 355$\cr 
.95&$\ \; \,1.355\ 439\ 724\ 384$&$\ \; \,1.355\ 483\ 452\ 108$&$\ \; \,1.355\ 515\ 734\ 412$
&$\ \; \,1.355\ 668\ 662\ 600$\cr 
1.05&$\ \; \,1.415\ 741\ 452\ 203$&$\ \; \,1.415\ 699\ 494\ 189$&$\ \; \,1.415\ 668\ 519\ 467$
&$\ \; \,1.415\ 521\ 797\ 633$\cr
3&$\ \; \,2.075\ 404\ 683\ 689$&$\ \; \,2.074\ 406\ 246\ 134$&$\ \; \,2.073\ 669\ 695\ 186$
&$\ \; \,2.070\ 187\ 162\ 577$\cr
4&$\ \; \,2.266\ 084\ 394\ 071$&$\ \; \,2.264\ 826\ 740\ 710$&$\ \; \,2.263\ 899\ 533\ 697$
&$\ \; \,2.259\ 524\ 751\ 387$\cr
\br 
\endalign 
\endtable 
\endinsert

In this letter, we deduce the universal amplitude of the free energy density
$A_0$ from the results of a finite-size scaling study (Bl\"ote and Nightingale
1982). A simple analytic expression is conjectured which is in excellent
agreement with the numerical results and reproduces the known exact value
$A_0\!=\!\pi/12$ in the Ising case, $q=2$ (Ferdinand and Fisher 1969).

Using the exact values for $B$ (Baxter 1973) and $f_0(L)$ for the three
largest strips shown in table 1, the universal amplitude $A_0(L)$ are obtained.
A three-point fit is used to estimate $A_0$, assuming a power law correction
to scaling
$$
A_0(L)=A_0+CL^{-y}\,.
\eqno(12)
$$
The results are shown in table 2.
\topinsert
\table{Universal amplitude $A_{0\ {\rm extrap}}$ obtained through a three-point fit of
$A_0(L)$ using equation (12).}[w] 
\align\L{#}&&\L{#}\cr  
\br  
&\centre{3}{$A_0(L)$}&\centre{1}{$A_{0\ {\rm extrap}}$}&&\cr  
\ns\ns
&\crule{3}&\crule{1}&&\cr
&\centre{4}{$L$}&&\cr
\ns\ns
&\crule{4}&&\cr
$q$&$\ \;\,9$&$\ \;\,10$&$\ \;\,11$&$\ \;\,\infty$&$y$&$\ \;\,C$\cr
\mr\cr
\ns\ns\ns
$1/64$&$-0.873\ 084\ 202\ 677$&$-0.872\ 315\ 316\ 000$&$-0.871\ 753\ 813\ 612$
&$-0.869\ 26$&$2.13$&$-0.41$\cr  
$1/16$&$-0.711\ 675\ 450\ 108$&$-0.711\ 006\ 032\ 400$&$-0.710\ 517\ 341\ 787$
&$-0.708\ 35$&$2.13$&$-0.36$\cr
$1/2$&$-0.234\ 795\ 928\ 284$&$-0.234\ 529\ 753\ 000$&$-0.234\ 335\ 551\ 934$
&$-0.233\ 47$&$2.14$&$-0.14$\cr
$0.95$&$-0.018\ 543\ 995\ 496$&$-0.018\ 521\ 049\ 200$&$-0.018\ 504\ 310\ 748$
&$-0.018\ 43$&$2.14$&$-0.01$\cr
$1.05$&$\ \;\,0.017\ 792\ 020\ 170$&$\ \;\,0.017\ 769\ 655\ 600$
&$\ \;\,0.017\ 753\ 341\ 914$&$\ \;\,0.017\ 68$&$2.14$&$\ \;\,0.01$\cr
$3$&$\ \;\,0.422\ 619\ 210\ 072$&$\ \;\,0.421\ 908\ 355\ 700$
&$\ \;\,0.421\ 386\ 445\ 689$&$\ \;\,0.419\ 00$&$2.08$&$\ \;\,0.35$\cr
$4$&$\ \;\,0.531\ 331\ 057\ 404$&$\ \;\,0.530\ 198\ 932\ 300$
&$\ \;\,0.529\ 348\ 659\ 510$&$\ \;\,0.524\ 94$&$1.85$&$\ \;\,0.37$\cr
\br 
\endalign 
\endtable
\endinsert
\goodbreak\midinsert 
\table{Universal amplitude $A_0$, $A_1$ and $A_2$ for the singular part ofthe
free energy levels of the $q$-state Potts model obtained from equations (16)
and (17) in the text. $A_{0\ {\rm extrap}}$ is presented for comparison.}[w]  
\align\L{#}&&\L{#}\cr   
\br  
$q$&$u$&$\ \;\,A_{0\ {\rm extrap}}$&$\ \;\,A_0$&$\ \;\,A_1$&$\ \;\,A_2$\cr   
\mr\cr
\ns\ns\ns
$0$&$1$&&$-\pi/3=$&$-\pi/3=$&$-13\pi/3=$\cr
&&&$-1.047\ 197\ 551$&$-1.047\ 197\ 551$&$-13.613\ 568\ 17$\cr
$1/64$&$0.960\ 185\ 314$&$-0.869\ 26$&$-0.869\ 154\ 052$&$-0.987\ 051\ 486$
&$-12.713\ 771\ 87$\cr   
$1/16$&$0.920\ 213\ 825$&$-0.708\ 35$&$-0.708\ 256\ 312$&$-0.931\ 130\ 284$
&$-11.881\ 819\ 72$\cr
$1/2$&$0.769\ 946\ 544$&$-0.233\ 47$&$-0.233\ 438\ 107$&$-0.753\ 415\ 910$
&$\ \,-9.274\ 429\ 021$\cr
$0.95$&$0.675\ 934\ 685$&$-0.018\ 43$&$-0.018\ 426\ 991$&$-0.662\ 744\ 760$
&$\ \,-7.971\ 364\ 109$\cr
$1$&$2/3$&&$\ \;\,0$&$-5\pi/24=$&$-5\pi/2=$\cr
&&&&$-0.654\ 498\ 470$&$\ \,-7.853\ 981\ 635$\cr
$1.05$&$0.657\ 551\ 944$&$\ \;\,0.017\ 68$&$\ \;\,0.017\ 677\ 994$
&$-0.646\ 499\ 628$&$\ \,-7.740\ 317\ 537$\cr
$2$&$1/2$&&$\ \;\,\pi/12=$&$-\pi/6=$&$-23\pi/12=$\cr
&&&$\ \;\,0.261\ 799\ 388$&$-0.523\ 598\ 776$&$\ \,-6.021\ 385\ 920$\cr
$3$&$1/3$&$\ \;\,0.419\ 00$&$\ \;\,2\pi/15=$&$-2\pi/15=$&$-22\pi/15=$\cr
&&&$\ \;\,0.418\ 879\ 021$&$-0.478\ 879\ 021$
&$\ \,-4.607\ 669\ 226$\cr
$4$&$0$&$\ \;\,0.524\ 94$&$\ \;\,\pi/6=$&$-\pi/12=$&$-5\pi/6=$\cr
&&&$\ \;\,0.523\ 598\ 776$&$-0.261\ 799\ 388$
&$\ \,-2.617\ 993\ 878$\cr
\br 
\endalign 
\endtable 
\endinsert

The exponents of the $q$-state Potts model are exactly known (den Nijs 1979,
Black and Emery 1981, Nienhuis \etal 1980, Pearson 1980, Wu 1982). Using the
variable $u$ defined as 
$$
0\leq u\equiv(2/\pi)\,\cos^{-1}\!(\sqrt{q}/2)\leq1
\eqno(13)
$$
one has
$$\eqalign{
y_t&=3(1-u)/(2-u)\qquad q\leq4\cr
y_h&=(3-u)(5-u)/4(2-u)\cr}
\eqno(14)
$$
and the decay exponents for the spin-spin and energy-energy correlation
functions are 
$$
\eqalign{
&\eta={1-u^2\over2(2-u)}\cr
&\eta_{ee}={2(1+u)\over 2-u}\,.\cr
}
\eqno(15)
$$
The numerical results obtained for $A_0$ may be fitted using the following
expression (table 3):
$$
A_0=\pi\,{(2-3u)(u+1)\over 6(2-u)}\,.
\eqno(16)
$$
The universal amplitudes for the second and third levels follow from equations
(11) and (15):
$$
\eqalign{
A_1&=-\pi\,{1+u\over 6(2-u)}\cr 
A_2&=-\pi\,{(1+u)(10+3u)\over 6(2-u)}\,.\cr
}
\eqno(17)
$$
\medskip
\noindent{\smallfonts\frenchspacing {\it Note added.} After this letter was
submitted for publication, we learnt that the result given in equation (16) had
been independently conjectured and derived from conformal invariance by Bl\"ote
\etal (1986) (see also Affleck 1986).}

\references
\refjl{Affleck I 1986}{\PRL}{56}{746}
\refjl{Baxter R J 1973}{J. Phys. C: Solid State Phys.}{6}{L445}
\refjl{Black J L and Emery V J 1981}{\PR\ {\rm B}}{23}{429}
\refjl{Bl\"ote H W J, Cardy J L and Nightingale M P 1986}{\PRL}{56}{742}
\refjl{Bl\"ote H W J and Nightingale M P 1982}{Physica}{112A}{405}
\refjl{Cardy J L 1984}{\JPA}{17}{L385}
\refjl{den Nijs MPM 1979}{\JPA}{12}{1857}
\refjl{Derrida B and de S\`eze J 1982}{J. Physique}{43}{475}
\refjl{Ferdinand A E and Fisher M E 1969}{\PR\ {\rm B}}{185}{832}
\refjl{Luck J M 1982}{\JPA}{15}{L169}
\refjl{Nienhuis B, Riedel E K and Schick M 1980}{\JPA}{13}{L189}
\refjl{Nightingale M P and Bl\"ote H W J 1983}{\JPA}{16}{L657}
\refjl{Pichard J L and Sarma G 1981}{J. Phys. C: Solid State Phys.}{14}{L617}
\refjl{Pearson R B 1980}{\PR\ {\rm B}}{22}{2579}
\refjl{Privman V and Fisher M E 1984}{\PR\ {\rm B}}{30}{322}
\refjl{Wu F Y 1982}{\RMP}{54}{235}
\vfill\eject\bye